\documentclass[12pt]{article}
\usepackage{latexsym}
\usepackage{amssymb}
\usepackage{amsmath}
\usepackage{amsthm}
\usepackage{graphicx}

\newtheorem{theorem}{Theorem:}
\newtheorem*{cor}{Corollary:}
\newcommand{\Epsilon}{\mathcal{E}}
\newcommand{\Tau}{\mathcal{T}}
\newcommand{\alf}{\alpha(r,t)}
\newcommand{\gam}{\gamma(r,t)}

\newcommand{\parl}{\shortparallel}
\newcommand{\lam}{\alpha}
\newcommand{\nw}{\gamma}

\begin{document}

\title{General solutions of Einstein's spherically symmetric gravitational equations with
junction conditions}
\author {{\small A. Das \footnote{e-mail: das@sfu.ca}} \\
\it{\small Department of Mathematics} \\ \it{\small Simon Fraser
University, Burnaby, British Columbia, Canada V5A 1S6}
 \and {\small A. DeBenedictis \footnote{e-mail: adebened@sfu.ca}} \\
\it{\small Department of Physics} \\ \it{\small Simon Fraser
University, Burnaby, British Columbia, Canada V5A 1S6} \and
{\small N. Tariq \footnote{e-mail: nessim@lums.edu.pk}} \\
\it{\small Department of Mathematics}\\ \it{\small Simon Fraser
University, Burnaby, British Columbia, Canada V5A 1S6} \\
\small{and} \\
\it{\small Department of Mathematics} \\ \it{\small Lahore
University of Management Sciences, D.H.A., Lahore Cantt.,
Pakistan}}
\date{{\small June 30, 2003}}
\maketitle
\begin{abstract}
\noindent Einstein's spherically symmetric interior gravitational
equations are investigated. Following Synge's procedure, the most
general solution of the equations is furnished in case
$T^{1}_{\;1}$ and $T^{4}_{\;4}$ are prescribed. The existence of
a total mass function, $M(r,t)$, is rigorously proved. Under
suitable restrictions on the total mass function, the
Schwarzschild mass $M(r,t)=m$, implicitly defines the boundary of
the spherical body as $r=B(t)$. Both Synge's junction conditions
as well as the continuity of the second fundamental form are
examined and solved in a general manner. The weak energy
conditions for an \emph{arbitrary boost} are also considered. The
most general solution of the spherically symmetric anisotropic
fluid model satisfying both junction conditions is furnished. In
the final section, various exotic solutions are explored using the
developed scheme including gravitational instantons, interior
$T$-domains and $D$-dimensional generalizations.
\end{abstract}

\section{Introduction}
\qquad As motivation, let us consider various solutions of a toy
model of the partial differential equation
\begin{equation}
\frac{\partial^{2}W(x,t)}{\partial x^{2}} - \frac{\partial^{2}
W(x,t)}{\partial t^{2}} =T_{0}. \label{eq:toyeqn}
\end{equation}
Here, $T_{0}$ is a prescribed constant. \emph{(i)} A particular
solution is provided by $W(x,t)=x-t+\frac{T_{0}}{4}(x^{2}-t^{2})$
which satisfies the initial value problem $W(x,0)=x+T_{0}x^{2}/4$
and $\frac{\partial W(x,t)}{\partial t}_{|t=0}=-1$. \emph{(ii)} A
class of general solutions of the same equation is given by
$W(x,t)=h(x+t)+T_{0}/4(x^{2}-t^{2})$, where $h$ is of class
$C^{2}$ but otherwise arbitrary. This class contains infinitely
many solutions but excludes infinitely many other solutions
including the solution in (i). \emph{(iii)} The most general
solution of this partial differential equation is furnished by
$W(x,t)=f(x-t)+g(x+t) +T_{0}/4(x^{2}-t^{2})$. Here both $f$ and
$g$ are of class $C^{2}$ and otherwise arbitrary. This class
contains \emph{all} possible (smooth) solutions of the equation.

\qquad Einstein's gravitational field equations,
$G^{i}_{\;j}+8\pi\frac{G}{c^{2}} T^{i}_{\;j}=0$, inside matter are
a system of second order, quasilinear, coupled partial
differential equations in four space-time variables. It is almost
impossible to obtain the most general solution for such a system
in case the $T^{i}_{\;j}$'s are prescribed. However, if the
space-time admits a group of motions or symmetry, then the
equations simplify considerably. In fact, in the arena of
spherical symmetry, using the curvature coordinates, Synge
\cite{ref:syngebook} obtained the most general solutions where
$T^{1}_{\;1}$ and $T^{4}_{\;4}$ are prescribed (the most logical
prescription from a physical perspective). The interior was
continuously patched to the exterior Schwarzschild metric across
the junction of the spherical material in the local sense.
However, the mathematical conditions assuring the existence of a
boundary were not derived. Moreover, satisfaction of Synge's own
junction conditions, $T^{i}_{\;j}n^{j}_{|\partial D}$ was not
completed.

\qquad In section 2, we write the spherically symmetric interior
equations in curvature coordinates. Then, we exhibit the most
general solution following Synge's prescriptions.

\qquad In section 3, we prove the mathematical existence of a
function, $M(r,t)$. Physically, this is the ``total mass'' of the
spherical body with coordinate radius $r$ at coordinate time $t$.
Under some reasonable assumptions, the implicit function theorem
\cite{ref:two} guarantees the existence of a solution to $r=B(t)$
for the equation $M(r,t)=m$, the Schwarzschild mass. The curve
$r=B(t)$ yields, in a natural way, the desired boundary for the
spherical body. It is important to note that this patching is
general and is therefore valid for junctions between various
interior layers (as in, for example, multi-layered stars) as well
as interior-vacuum patching.

\qquad In section 4, we obtain necessary and sufficient conditions
for the satisfaction of Synge's junction conditions
\cite{ref:syngebook} across the junction. Moreover, we also
investigate the Israel-Sen-Lanczos-Darmois (ISLD) junction
conditions \cite{ref:three} across the junction and obtain general
solutions of the problem.

\qquad In the next section, we examine the weak energy conditions
\cite{ref:four} thoroughly for the spherically symmetric scenario.
We obtain the general solution of the inequalities in terms of
four arbitrary slack functions.

\qquad In section 6, the class of spherically symmetric
$\left[T^{i}_{\;j}\right]$ with real eigenvalues is critically
studied. As a particular application, the anisotropic fluid model
(which contains the perfect fluid as a special case) is explored
exhaustively. Theorems are proved on the most general solution of
the corresponding field equations with \emph{both} junction
conditions of Synge and those of ISLD. Other special examples
(black holes etc.) are also treated.

\qquad In the last section, exotic spherically symmetric
solutions and their relation to the proposed scheme are explored.
Signature changing metrics as well as the Euclidean gravitational
instantons \cite{ref:five} are furnished. Next, $T$-domain
\cite{ref:six} equations and general solutions are provided. A
special class of $T$-domain solutions yields the so called
eternal black holes. Another special class of $T$-domain
solutions involve complex eigenvalues of the stress-energy
tensor. Such examples were already found in exotic black holes
\cite{ref:eight}. Finally, we give motivation for, and briefly
investigate, spherically symmetric interior equations in
arbitrary dimension $D\geq 3$. The corresponding general solution
is provided \cite{ref:nine}.

\section{Solution of the spherically symmetric field equations}
\qquad We adopt notations and conventions from Synge's book
\cite{ref:syngebook}, except that covariant derivatives are
denoted by $\nabla_{k}$. Physical units are chosen so that $c=1$
and $\kappa:=8\pi G$.

\qquad Einstein's gravitational equations are furnished by:
\begin{subequations}
\begin{align}
&\mathcal{E}_{ij}:=G_{ij}+\kappa T_{ij} =0, \label{eq:einst} \\
&\Tau^{i}:=\nabla_{j}T^{ij} =0, \label{eq:cons} \\
&\nabla_{j}\mathcal{E}^{ij}-\kappa\Tau^{i} \equiv 0.
\label{eq:bianchi}
\end{align}
\end{subequations}
It is assumed that the metric functions, $g_{ij}(x)$, are of
class $C^{3}$ and the functions $T_{ij}(x)$ are of class $C^{1}$.

\qquad A spherically symmetric metric, in the curvature coordinate
chart, and the natural orthonormal tetrad are characterized by
\begin{eqnarray}
ds^{2}&=&e^{\alf}\,dr^{2}+r^{2}\left[d\theta^{2}+\sin^{2}\theta
\,d\phi^{2}\right] -e^{\gam}\,dt^{2},
\label{eq:metric} \\
e^{i}_{(1)}&=&e^{-\alf/2}\delta^{i}_{(1)}, \;
e^{i}_{(2)}=r^{-1}\delta^{i}_{(2)}, \; e^{i}_{(3)}=(r
\sin\theta)^{-1} \delta^{i}_{(3)}, \;
e^{i}_{(4)}=e^{-\gam/2}\delta^{i}_{(4)}. \nonumber
\end{eqnarray}
Non-trivial equations and identities from
(\ref{eq:einst}-\ref{eq:metric}) are provided by:
\begin{subequations}
\begin{align}
\Epsilon^{1}_{\;1}=&r^{-2}\left[1-e^{-\alpha}(1+r\gamma_{,1})\right]
+\kappa T^{1}_{\;1} =0, \label{eq:einstone}\\
\Epsilon^{2}_{\;2}\equiv
\Epsilon^{3}_{\;3}=&\frac{1}{2}e^{-\alpha} \left[-\gamma_{,11}
+\frac{1}{2r}\left(r\gamma_{,1}+2\right)(\alpha-\gamma)_{,1}
\right] \nonumber \\
&+\frac{1}{2}e^{-\gamma}\left[\alpha_{,44} +\frac{1}{2}
\alpha_{,4}\left(\alpha-\gamma\right)_{,4} \right] +\kappa
T^{2}_{\,2} =0, \label{eq:einsttwo} \\
\Epsilon^{1}_{\;4}=&\frac{1}{r}\left(e^{-\alpha}\right)_{,4}
+\kappa T^{1}_{\;4} =0, \label{eq:einstthree} \\
\Epsilon^{4}_{\;4}=&\frac{1}{r^{2}}\left[1-(re^{-\alpha})_{,1}\right]
+\kappa T^{4}_{\;4} =0, \label{eq:einstfour} \\
\Tau_{1}=&T^{1}_{\;1,1}+T^{4}_{\;1,4}+\frac{2}{r}
\left[1+\frac{r}{4}\gamma_{,1}\right]T^{1}_{\;1}+\frac{1}{2}
\left(\alpha+\gamma\right)_{,4} T^{4}_{\;1} \nonumber \\
&-\frac{1}{2}\gamma_{,1}T^{4}_{\;4} -\frac{2}{r}T^{2}_{\;2} =0, \label{eq:cons1} \\
\Tau_{4}=&T^{1}_{\;4,1}+T^{4}_{\;4,4}+\frac{2}{r}
\left[1+\frac{r}{4} (\alpha+\gamma)_{,1}\right] T^{1}_{\;4}
+\frac{1}{2} \alpha_{,4}\left(T^{4}_{\;4}-T^{1}_{\;1}\right) =0,
\label{eq:cons2}\\
\Epsilon^{1}_{\;1,1}+&\Epsilon^{4}_{\;1,4} +\frac{2}{r}
\left[1+\frac{r}{4}\gamma_{,1}\right] \Epsilon^{1}_{\;1}
+\frac{1}{2}\left(\alpha+\gamma\right)_{,4}\Epsilon^{4}_{\;1}
\nonumber \\
&-\frac{1}{2}\gamma_{,1}\Epsilon^{4}_{\;4}
-\frac{2}{r}\Epsilon^{2}_{\;2} -\kappa \Tau_{1} \equiv 0,\label{eq:bianchione} \\
\Epsilon^{1}_{\;4,1}+&\Epsilon^{4}_{\;4,4}+\frac{2}{r}
\left[1+\frac{r}{4}\left(\alpha+\gamma\right)_{,1}
\right]\Epsilon^{1}_{\;4} +\frac{1}{2}
\alpha_{,4}\left(\Epsilon^{4}_{\;4}-\Epsilon^{1}_{\;1}\right)
-\kappa\Tau_{4} \equiv 0. \label{eq:bianchitwo}
\end{align}
\end{subequations}

\qquad We study and solve these equations in a two dimensional
domain given by:
\begin{equation}
D:=\left\{(r,t):\; 0<r<B(t),\;t_{1} < t <t_{2}\right\}.
\label{eq:domain}
\end{equation}
Note that one may relax the restriction to the domain $r_{0} < r
< B(t)$ in which case radial integrals in the following should
possess the lower limit of $r_{0}$. In such a case, an inner
boundary will exist at $r_{0}$ and the junction conditions
discussed later should be applied to the inner boundary as well.
The outer boundary curve, $r=B(t)$, will be explicitly determined
later (see fig. 1).

\qquad Synge's strategy of solving the field equations is the
following:
\begin{itemize}
\item Prescribe $T^{4}_{\;4}\equiv T^{(4)}_{\;(4)}$ from desirable physical properties and solve
$\Epsilon^{4}_{\;4}=0$ to obtain $e^{-\alpha}=g_{11}$.
\item Prescribe $T^{1}_{\;1}\equiv T^{(1)}_{\;(1)}$ or relate it
to $T^{4}_{\;4}$ by an equation of state and solve
$\Epsilon^{1}_{\;1}-\Epsilon^{4}_{\;4}=0$ to obtain
$e^{\gamma}=-g_{44}$.
\item Define $T^{1}_{\;4}$ by the equation $\Epsilon^{1}_{\;4}=0$.
\item Define $T^{2}_{\;2}\equiv T^{(2)}_{\;(2)}$ by the
conservation equation $\Tau_{1}=0$.
\item By the preceding step, the identity (\ref{eq:bianchione})
implies that $\Epsilon^{2}_{\;2}=0$.
\item By the identity (\ref{eq:bianchitwo}), the conservation
equation $\Tau_{4}=0$ is satisfied.
\end{itemize}
At this stage, \emph{all the field equations, conservation laws
and identities are satisfied}. One may impose further
restrictions to the above scheme. For example, in the case of the
perfect fluid, the conservation equation (\ref{eq:cons1}) becomes
a differential equation for the pressure (or the energy density,
if an equation of state exists), which must be solved. As well,
one may require that further equations, such as matter field
equations, need to be satisfied.

\qquad Regardless of the variants on the above scheme, all
solutions must satisfy the following most general solution
yielded by:
\begin{subequations}
\begin{align}
e^{-\alf}=&1-\frac{\kappa}{r} \left[f^{2}(t)-\int_{0_{+}}^{r}
T^{4}_{\;4}(x,t)\,x^{2}\,dx\right], \label{eq:ealpha} \\
e^{\gam}=&e^{-\alf} \left\{\exp\left[h(t)+\kappa \int_{0_{+}}^{r}
\left[T^{1}_{\;1}(x,t)-T^{4}_{\;4}(x,t)\right] e^{\alpha(x,t)}
x\,dx\right]\right\}, \label{eq:egamma}\\
T^{1}_{\;4}(r,t):=&\frac{1}{r^{2}} \left[ 2f(t)\dot{f}(t)
-\int_{0_{+}}^{r}T^{4}_{\;4,4} x^{2}\,dx\right], \label{eq:energyflux} \\
T^{2}_{\;2}\equiv T^{3}_{\;3}
:=&\frac{r}{2}\left[T^{1}_{\;1,1}+T^{4}_{\;1,4}\right]
+\left[1+\frac{r}{4}\gamma_{,1}\right]T^{1}_{\;1}
+\frac{r}{4}\left(\alpha+\gamma\right)_{,4}T^{4}_{\,1}
-\frac{r}{4}\gamma_{,1}T^{4}_{\;4}, \label{eq:pressure}
\end{align}
\end{subequations}
with $\dot{f}(t):=\frac{df(t)}{dt}$. Here, $f(t)$ and $h(t)$ are
\emph{two arbitrary functions of integration} which are of class
$C^{3}$. Synge \cite{ref:syngebook} set $f^{2}(t)\equiv 0$ to
avoid a singularity at the center. However, this function may be
important in certain cases such as the study of wormholes. The
function $h(t)$ was absorbed by a transformation of the time
coordinate though this is not always possible \cite{ref:nine} -
\cite{ref:otherbirks}. We retain these functions for generality
and to satisfy junction conditions later.

\section{Conservation equations, the total mass function and the boundary}
\qquad We notice from the equations (\ref{eq:einstthree}) and
(\ref{eq:einstfour}) the existence of two \emph{additional}
differential identities:
\begin{subequations}
\begin{align}
\left(r^{2}G^{1}_{\;4}\right)_{,1}
+\left(r^{2}G^{4}_{\;4}\right)_{,4} &\equiv 0, \\
(\alpha+\gamma)_{,1}G^{1}_{\;4}
+\alpha_{,4}\left(G^{4}_{\;4}-G^{1}_{\;1}\right) &\equiv 0.
\end{align}
\end{subequations}
However, because of $\nabla_{k}G^{k}_{\;4}\equiv 0$, only one of
the above additional identities is independent. Therefore, there
must exist additional conservation equations
\begin{subequations}
\begin{align}
\Tau_{4a}:=&\left(r^{2}T^{1}_{\;4}\right)_{,1}
+\left(r^{2}T^{4}_{\;4}\right)_{,4} = 0, \label{eq:newcons1} \\
\Tau_{4b}:=&(\alpha+\gamma)_{,1}T^{1}_{\;4}
+\alpha_{,4}\left(T^{4}_{\;4}-T^{1}_{\;1}\right) = 0,
\label{eq:newcons2} \\
d\left[\frac{\kappa}{2}\right. & \left. r^{2} T^{1}_{\;4}(r,t)\,
dt
-\frac{\kappa}{2} r^{2} T^{4}_{\;4}(r,t)\, dr \right]=0, \\
\Tau_{4}\equiv& \frac{1}{r^2}\left[\Tau_{4a}+\Tau_{4b}\right].
\end{align}
\end{subequations}
The first of these equations has a simple physical
interpretation. Integrating over a sphere, the equation relates
the rate of change of energy in a sphere of radius $r$ to the
total energy flux entering or leaving the boundary of the sphere.
The equation (\ref{eq:newcons1}) in the star-shaped domain $D$
guarantees, by the converse Poincar\'{e} lemma \cite{ref:ten} the
existence of a function $M(r,t)$ of class at least $C^{2}$ such
that
\begin{subequations}
\begin{align}
dM(r,t)=&\frac{\kappa r^{2}}{2}
\left[T^{1}_{\;4}(r,t)\,dt-T^{4}_{\;4}(r,t)\,dr \right],
\label{eq:massdifferential} \\
M_{,1}=&-\frac{\kappa}{2}r^{2}T^{4}_{\;4}(r,t), \label{eq:massderiv1} \\
M_{,4}=&\frac{\kappa}{2}r^{2}T^{1}_{\;4}(r,t).
\label{eq:massderiv2}
\end{align}
\end{subequations}
From the equations (\ref{eq:einstfour}), (\ref{eq:ealpha}) and
(\ref{eq:energyflux}), we conclude that
\begin{subequations}
\begin{align}
2M(r,t)=&\kappa f^{2}(t)-\kappa\int_{0_{+}}^{r} T^{4}_{\;4}(x,t)
x^{2}\, dx, \label{eq:2M} \\
\lim_{r\rightarrow 0^{+}}M(r,t)=&\frac{\kappa}{2}f^{2}(t) , \label{eq:limM} \\
e^{-\alf}=&1-\frac{2M(r,t)}{r} , \label{eq:eminusalf}\\
e^{\gam}=&\left[1-\frac{2M(r,t)}{r}\right]
\exp\left[h(t)+\chi(r,t)\right], \label{eq:etogamma} \\
\chi(r,t):=&\kappa\int_{0}^{r}\left[\frac{T^{1}_{\;1}(x,t)-T^{4}_{\;4}(x,t)}
{x-2M(x,t)}\right] x^{2}\, dx. \label{eq:chi}
\end{align}
\end{subequations}
We tacitly assume that $r-2M(r,t)\neq0$ in $D$. The physical
interpretation of $M(r,t)$ is the ``total mass'' contained in the
spherical volume of ``radius'' $r$ and at ``time'' $t$.

\qquad Next we wish to study the level curves of the function
$M(r,t)$. For the existence of such curves we state the following
version of the implicit function theorem \cite{ref:two}.

{\theorem{Let $M(r,t)$ be a function of at least class $C^{1}$ in
$D$ such that for a point $(r_{0},t_{0})$ in $D$, the function
$M(r_{0},t_{0})=c$, a constant. Suppose that
$M_{,1|(r_{0},t_{0})} \neq0$. Then there exists a function
$B(t;c)$ of class at least $C^{1}$ in the neighborhood of
$(r_{0},t_{0})$ such that $r=B(t;c)$ is a solution of $M(r,t)=c$
in that neighborhood with $r_{0}=B(t_{0};c)$.}}

\qquad The boundary curve $r=B(t)$ of the spherical body in the
definition (\ref{eq:domain}) is defined by the following:
\begin{eqnarray}
B(t)&:=&\lim_{c\rightarrow m_{-}}B(t;c), \label{eq:boundaryc} \\
\partial D&:=&\left\{(r,t):r=B(t),\;\; t_{1}<t<t_{2}\right\},
\nonumber
\end{eqnarray}
(see figure 1). Here, $m>0$ physically represents the total
Schwarzschild mass of the body. It is assumed that $M>0$, $M_{,1}
>0$ or $T^{4}_{\;4} < 0$ in $D$.

\begin{figure}[ht]
\begin{center}
\includegraphics[bb=113 381 441 563, scale=0.7, keepaspectratio=true]{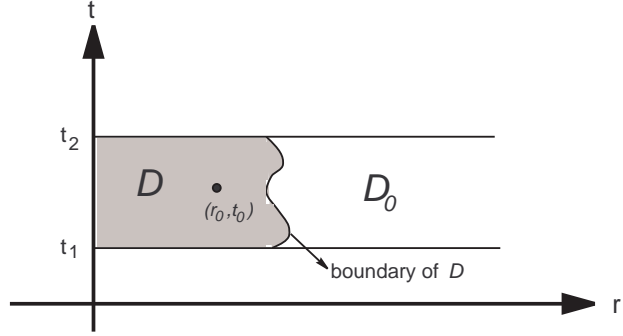}
\caption{{\small The considered domain with boundary, $\partial
D$, which separates the interior domain $D$ and the vacuum domain
$D_{0}$.}} \label{fig1}
\end{center}
\end{figure}

It is clear from the implicit definition of the boundary curve,
$\partial D:\;\;r=B(t)$ that
\begin{eqnarray}
M(B(t),t)&\equiv& m, \nonumber \\
\dot{B}(t)&=&-\left[\frac{M_{,4}}{M_{,1}}\right]_{|\partial D}
=\left[\frac{T^{1}_{\;4}}{T^{4}_{\;4}}\right]_{|\partial D}.
\label{eq:bdotmass}
\end{eqnarray}
The spherical body collapses in case $M_{,4}(r,t)>0$ and expands
in case $M_{,4}(r,t)<0$.

\qquad In case the measurable speed of the boundary is less than
the speed of light, we must have:
\begin{eqnarray}
e^{\alpha-\gamma}_{|\partial D}\left[\dot{B}(t)\right]^{2} &<&1,
\nonumber \\
\left[e^{-\alpha}(M_{,1})^{2}-e^{-\gamma}(M_{,4})^{2}\right]_{|\partial
D} &>& 0, \label{eq:ineq} \\
\left[\frac{T^{1}_{\;4}}{T^{4}_{\;4}}\right]^{2}_{|\partial D} &<&
\left[1-\frac{2m}{B(t)}\right]^{2}\exp\left[h(t)+\chi(B(t),t)\right],
\nonumber \\
B(t)&\neq& 2m. \nonumber
\end{eqnarray}

\qquad The interior domain $D$, the boundary $\partial D$ and the
exterior (vacuum) domain $D_{0}$ are explained in the equations
(\ref{eq:ealpha}),(\ref{eq:egamma}) and (\ref{eq:boundaryc}) and
in the figure 1. Following Synge, we shall now match continuously
the interior metric to the exterior metric (transformable to the
Schwarzschild chart). We must use the equations
(\ref{eq:etogamma}), (\ref{eq:chi}) and (\ref{eq:boundaryc}) to
arrive at:

\begin{equation}
 g^{11}(r,t)=e^{-\alf}=\left\{
\begin{array}{lll}
1-\frac{2M(r,t)}{r} & \mbox{ for }  & 0<r<B(t),\;\;t_{1}< t < t_{2}, \\
1-\frac{2m}{B(t)} & \mbox{ for }&
r=B(t),\;\;t_{1}< t < t_{2} ,  \\
1-\frac{2m}{r} & \mbox{ for }& B(t) < r < \infty, \;\; t_{1}< t <
t_{2},
\end{array}
\right. \label{eq:g11}
\end{equation}

\begin{equation}
-g_{44}(r,t)=e^{\gam}=\left\{
\begin{array}{lll}
\left[1-\frac{2M(r,t)}{r}\right]\exp\left[h(t)+\chi(r,t)\right] & \mbox{ for }  & 0<r<B(t),\;\;t_{1}< t < t_{2}, \\
\left[1-\frac{2m}{B(t)}\right]\exp\left[h(t)+\chi(B(t),t)\right] &
\mbox{ for }&
r=B(t),\;\;t_{1}< t < t_{2} ,  \\
\left[1-\frac{2m}{r}\right]\exp\left[h(t)+\chi(B(t),t)\right] &
\mbox{ for }& B(t) < r < \infty, \;\; t_{1}< t < t_{2}.
\end{array}
\right. \label{eq:g44}
\end{equation}
The exterior metric can be easily transformed to the
Schwarzschild coordinates. We shall next investigate both Synge's
\cite{ref:syngebook} and ISLD's \cite{ref:three} junction
conditions.

\section{Junction conditions}
\subsection{Synge's junction condition}
\qquad Synge's junction conditions read
\begin{eqnarray}
T^{i}_{\;j}n_{i|\partial D}&=&0, \label{eq:syngecond} \\
n_{i}n^{i} &=&1 \nonumber,
\end{eqnarray}
with $n^{i}$ a unit normal at the boundary. In the present case,
with the help of (\ref{eq:ineq}) we can write for the relevant
normal components:
\begin{eqnarray}
n_{1}=\frac{M_{,1}}{\sqrt{e^{-\alpha}(M_{,1})^{2}- 
e^{-\gamma}(M_{,4})^{2}}}|_{\partial
D} , \\
n_{4}=\frac{M_{,4}} 
{\sqrt{e^{-\alpha}(M_{,1})^{2}-e^{-\gamma}(M_{,4})^{2}}}|_{\partial
D}. \nonumber
\end{eqnarray}
The equations (\ref{eq:syngecond}) reduce to
\begin{subequations}
\begin{align}
\left[T^{1}_{\;1}M_{,1} +T^{4}_{\;1}M_{,4}\right]_{|\partial D}
\label{eq:syngecond1}
&=0, \\
\left[T^{1}_{\;4}M_{,1} +T^{4}_{\;4}M_{,4}\right]_{|\partial D}
&=0. \label{eq:syngecond2}
\end{align}
\end{subequations}
By the equations (\ref{eq:massderiv1}) and (\ref{eq:massderiv2}),
the junction condition (\ref{eq:syngecond2}) is \emph{identically
satisfied}. Moreover, the other junction condition
(\ref{eq:syngecond1}) yields:
\begin{equation}
\left|
\begin{array}{ll}
T^{1}_{\;1} & T^{1}_{\;4} \\
T^{4}_{\;1} & T^{4}_{\;4}
\end{array}\right|_{\partial D} =0 \nonumber
\end{equation}
or,
\begin{equation}
\left[e^{\gamma-\alpha}T^{1}_{\;1} M_{,1}-
T^{1}_{\;4}M_{,4}\right]_{|\partial D} =0. \label{eq:determinant}
\end{equation}
There exist two possible cases here. In case the boundary is
static, we must have from (\ref{eq:bdotmass}) and (\ref{eq:determinant})
\begin{eqnarray}
\dot{B}(t)&\equiv& 0, \nonumber \\
T^{1}_{\;1}T^{4}_{\;4\,|\partial D}&\equiv& 0, \\
&h(t)& \mbox{is an arbitrary function} \label{eq:bdotzero}.
\end{eqnarray}
This case does \emph{not} imply that the interior metric is
necessarily static.

\qquad In case the boundary is non-static, we obtain from
(\ref{eq:massderiv1}), (\ref{eq:massderiv2}), (\ref{eq:bdotmass}) and
(\ref{eq:determinant})
\begin{eqnarray}
\dot{B}(t)&\neq&0, \nonumber \\
T^{1}_{\;1}T^{4}_{\;4 |\partial D} &<&0, \nonumber \\
e^{h(t)}&=&\left[1-\frac{2m}{B(t)}\right]^{-2}
\exp\left[-\chi(B(t),t)\right]\left[\dot{B}(t)\right]^{2}
\left|\frac{T^{4}_{\;4}}{T^{1}_{\;1}}\right|_{\;|\partial D} > 0.
\label{eq:eht}
\end{eqnarray}
Thus, the function $h(t)$, which originated as an arbitrary
function of integration, can be utilized to
satisfy the junction conditions.

\subsection{Israel-Sen-Lanczos-Darmois junction condition}
\qquad Next we consider the ISLD junction conditions. Namely, we
consider the continuity of the second fundamental form at the
junction. For this purpose, the three-dimensional metric for the
hyper-surface corresponding to $\partial D \times S^{2}$ is
obtained from (\ref{eq:g44}) as
\begin{eqnarray}
d\sigma^{2} :=ds^{2}_{|r=B(t)}&=&B^{2}(t)\left[d\theta^{2}+
\sin^{2}\theta\,d\phi^{2}\right] \nonumber \\
&&-\left\{\left[1- \frac{2m}{B(t)}\right]\exp\left[h(t)
+\chi(B(t),t)\right]-\left[\dot{B}(t)\right]^{2} \right\}dt^{2}.
\label{eq:boundline}
\end{eqnarray}
The extrinsic curvature components \cite{ref:eleven} calculated
from the interior and exterior metrics are the following:
\begin{eqnarray}
K_{\theta\theta}^{\pm}&=& \lim_{\delta\rightarrow 0_{+}}
\left\{-\frac{re^{-\alpha}}{\sqrt{e^{-\alpha}-e^{-\gamma}
\left[\dot{B}(t)\right]^{2}}} \right\}_{|r=B(t)\pm\delta},
\nonumber \\
K^{\pm}_{\phi\phi}&=&\sin^{2}\theta\,K^{\pm}_{\theta\theta}, \\
2K^{\pm}_{tt}&=&\lim_{\delta\rightarrow 0_{+}} \left\{
\frac{1}{\sqrt{e^{-\alpha}-e^{-\gamma}
\left[\dot{B}(t)\right]^{2}}}
\left[2\ddot{B}(t)+e^{\gamma-\alpha} \gamma_{,1}
+\dot{B}(t)\left(2\alpha-\gamma\right)_{,4} \right.\right.
\nonumber \\
&&\left.\left.+\left[\dot{B}(t)\right]^{2}
\left(\alpha-2\gamma\right)_{,1}
-\left[\dot{B}(t)\right]^{3}e^{\alpha-\gamma}
\alpha_{,4}\right]\right\}_{|r=B(t)\pm\delta}.  \nonumber
\end{eqnarray}
It is clear from (\ref{eq:g11}), (\ref{eq:g44}) and
(\ref{eq:boundline}) that
\begin{eqnarray}
K^{-}_{\theta\theta} - K^{+}_{\theta\theta} &\equiv& 0, \\
K^{-}_{\phi\phi} - K^{+}_{\phi\phi} &\equiv& 0.
\end{eqnarray}
To show the continuity of $K_{tt}$ across $\partial D$, we
consider the function
\begin{equation}
2L^{\pm}:=\lim_{\delta\rightarrow 0_{+}} \left\{e^{\gamma-\alpha}
\gamma_{,1} +\dot{B}(t)(2\alpha-\gamma)_{,4}
+\left[\dot{B}(t)\right]^{2}(\alpha-2\gamma)_{,1}
-\left[\dot{B}(t)\right]^{3}e^{\alpha-\gamma}\alpha_{,4}\right\}_{r=B(t)\pm\delta}
.
\end{equation}
The continuity of $L$ across $\partial D$ implies, from
(\ref{eq:g44}) and (\ref{eq:einstone} - \ref{eq:cons2}), (after a long calculation) the
following algebraic equation:
\begin{eqnarray}
0&\equiv&
2\kappa^{-1}\left[1-\frac{2m}{B(t)}\right]e^{h(t)+\chi(B(t),t)}
\left[L^{-}-L^{+}\right] \nonumber \\
&=&U(t)e^{2h(t)}+ V(t)e^{h(t)} + W(t).
\end{eqnarray}
Here,
\begin{eqnarray}
U(t)&:=&B(t)\left[1-\frac{2m}{B(t)}\right]^{2}
e^{2\chi(B(t),t)}T^{1}_{\;1|r=B(t)},
\nonumber \\
V(t)&:=&-B(t)\dot{B}^{2}(t)\left\{e^{\chi(B(t),t)}\left(T^{1}_{\;1}
-T^{4}_{\;4}\right)\right\}_{| r=B(t)}, \\
W(t)&:=& -B(t)\left[\dot{B}(t)\right]^{4}
\left[1-\frac{2m}{B(t)}\right]^{-2}T^{4}_{\;4|r=B(t)}. \nonumber
\end{eqnarray}
Analyzing the above quadratic (or possibly linear) equation for
$e^{h(t)}$, we obtain the following solutions:

\begin{subequations}
\begin{align}
\mbox{Case I:}&\;\; T^{1}_{\;1|\partial D}=\dot{B}(t)\, T^{4}_{\;4|\partial D}
\equiv 0 \; \mbox{ and }\; h(t) \; \mbox{ is arbitrary;} \label{eq:israeli} \\
\mbox{Case II:}&\;\; 
e^{h(t)}=
\left[\dot{B}(t)\right]^{2}e^{-\chi(B(t),t)}
\left[1-\frac{2m}{B(t)}\right]^{-2}\;\;\; \mbox{ for }\;\;\;
\left\{
\begin{array}{lll}
i)\; T^{1}_{\;1|\partial D}\equiv 0, \;\; \dot{B}(t)T^{4}_{\;4|\partial
D} \neq 0; \\
ii) \; \left(T^{1}_{\;1}-T^{4}_{\;4} \right)_{\partial D} =0 ; \\
iii) \; T^{1}_{\;1|\partial D}\neq 0, \;\; T^{4}_{\;4|\partial D} =0;
\end{array}
\right. \label{eq:israelii} \\
\mbox{Case III:}& \;\; T^{1}_{\;1|\partial D} > 0, \;\;
T^{4}_{\;4|\partial D} <0, \;\; \dot{B}(t) \neq 0, \nonumber \\
&\mbox{ and }\;\; e^{h(t)}=-\left[\dot{B}(t)\right]^{2}
e^{-\chi(B(t),t)} \left[1-\frac{2m}{B(t)}\right]^{-2}
\left|\frac{T^{4}_{\;4}}{T^{1}_{\;1}}\right|_{|\partial D}.
\label{eq:israeliii} \\
\nonumber
\end{align}
\end{subequations}

It is clear that Synge's conditions (\ref{eq:bdotzero}) and
(\ref{eq:eht}) satisfy the ISLD conditions (\ref{eq:israeli}) and
(\ref{eq:israeliii}). Case II represents a possible mathematical extension of
the ISLD junction conditions to a non-time-like boundary.

\section{Weak energy conditions}
\qquad Next the weak energy conditions in spherical symmetry are
studied. We consider an observer with an \emph{arbitrary boost}
which, to our knowledge, has not been calculated before.

\qquad In terms of the orthonormal components (denoted by indices
in parentheses), the weak energy conditions \cite{ref:four} can be
stated as
\begin{equation}
T_{(a)(b)}u^{(a)}u^{(b)} \geq 0 \label{eq:WEC}
\end{equation}
for every time-like vector $u^{(a)}$ satisfying
\begin{equation}
\left[u^{(1)}\right]^{2}+ \left[u^{(2)}\right]^{2}+
\left[u^{(3)}\right]^{2} -\left[u^{(4)}\right]^{2} =-1 ,
\label{eq:timecond}
\end{equation}
with $u^{(4)}>0$ as dictated by reasonable physics.

\qquad The general solution of the above non-linear algebraic
equation (\ref{eq:timecond}) is given by:
\begin{eqnarray}
u^{(1)}&=&\sinh\beta\cos\theta, \;
u^{(2)}=\sinh\beta\sin\theta\cos\phi,\;u^{(3)}=\sinh\beta\sin\theta\sin\phi,
\; u^{(4)}=\cosh\beta, \nonumber \\
\beta &\in& \mathbb{R}, \;\; \theta\in(0,\,\pi), \;\; \phi \in
(-\pi,\,\pi). \label{eq:velocities}
\end{eqnarray}
For spherical symmetry, choosing the orthonormal basis of
(\ref{eq:metric}), the inequality (\ref{eq:WEC}) together with
equations (\ref{eq:velocities}) yield:
\begin{eqnarray}
&&\left[T_{(1)(1)}-T_{(2)(2)}\right]x^{2}y^{2} + T_{(2)(2)}x^{2}
+2T_{(1)(4)}xy +T_{(4)(4)} \geq 0, \label{eq:tineq} \\
&&x:=\tanh\beta,\;\; y:=\cos\theta. \nonumber
\end{eqnarray}
Analyzing the inequality (\ref{eq:tineq}) for all $(x,y) \in
[-1,1]\times [-1,1]$, we conclude, after much calculation, that
either
\begin{subequations}
\begin{align}
(i)&\;\;T_{(1)(1)}\equiv T_{(2)(2)}, \; T_{(1)(4)}\equiv 0,\;
T_{(4)(4)}
\geq 0,\; T_{(4)(4)}+T_{(1)(1)} \geq 0, \label{eq:econd1} \\
 \mbox{or else} \nonumber \\
(ii)& \;\; T_{(1)(1)} > T_{(2)(2)},\; \left(T_{(1)(4)}\right)^{2}
\leq
T_{(4)(4)}\left[T_{(1)(1)}-T_{(2)(2)}\right], \nonumber \\
& \; \left(T_{(1)(4)}\right)^{2} \leq
\left[T_{(1)(1)}-T_{(2)(2)}\right]\left[T_{(4)(4)}+
T_{(2)(2)}\right] \label{eq:econd2}.
\end{align}
\end{subequations}
We can solve the inequalities (\ref{eq:econd1}) (\ref{eq:econd2})
by utilizing slack functions:
\begin{subequations}
\begin{align}
T^{4}_{\;4}=&T^{(4)}_{\;(4)}=-E^{2}(r,t), \label{eq:slack1}\\
T^{2}_{\;2}=&T^{(2)}_{\;(2)}=P^{2}(r,t)-\left[
E(r,t)\sin Q(r,t)\right]^{2},\label{eq:slack2} \\
T^{1}_{\;1}=&T^{(1)}_{\;(1)}=P^{2}(r,t)-\left[
E(r,t)\sin Q(r,t)\right]^{2}+H^{2}(r,t), \label{eq:slack3} \\
e^{(\alpha-\gamma)/2}T^{1}_{\;4}=&T_{(1)(4)}= -H(r,t)E(r,t)\cos
Q(r,t) \label{eq:slack4}
\end{align}
\end{subequations}
Here, the slack functions $E(r,t),\;P(r,t),\;Q(r,t),\;H(r,t)$ are
of class $C^{1}$ but otherwise arbitrary.

\section{Real eigenvalues of $\left[T^{i}_{\;j}\right]$ and anisotropic fluid models}
\qquad  First, we analyze and solve the problem of a spherically
symmetric $T^{i}_{\;j}$ possessing real eigenvalues. Recall that
the eigenvalue problem for $T^{i}_{\;j}$ is given by:
\begin{equation}
T^{i}_{\;j}\,E^{j}_{(a)}=\lambda_{(a)}E^{i}_{(a)}\; .
\label{eq:eigval}
\end{equation}
In the spherically symmetric case,
$T^{1}_{\;2}=T^{1}_{\;3}=T^{2}_{\;4}=T^{3}_{\;4} \equiv 0, \;
T^{2}_{\;2}\equiv T^{3}_{\;3}$. Therefore, the eigenvalues of
$T^{i}_{\;j}$ are given by:
\begin{eqnarray}
\lambda_{(2)}&\equiv& \lambda_{(3)}=T^{2}_{\;2}, \label{eq:eigvals} \\
2\lambda_{(1)}&=&T^{1}_{\;1}+T^{4}_{\;4}+\sqrt{\Delta}, \nonumber \\
2\lambda_{(4)}&=&T^{1}_{\;1}+T^{4}_{\;4}-\sqrt{\Delta},
\nonumber \\
\Delta:&=&\left(T^{1}_{\;1}-T^{4}_{\;4}\right)^{2}-e^{\alpha-\gamma}
\left(T^{1}_{\;4}\right)^{2} . \nonumber
\end{eqnarray}
It is clear that $\Delta < 0$ will imply complex eigenvalues. We
restrict ourselves in this section to the case where the
stress-energy tensor possesses real eigenvalues ($\Delta \geq 0$).

\qquad In a static model, $T^{1}_{\;4}\equiv 0$ and $\Delta \geq
0$ is automatically valid. In case of the weak energy conditions
in (\ref{eq:econd1}), (\ref{eq:econd2}) and the corresponding solutions
in (\ref{eq:slack1} - \ref{eq:slack4}),
\begin{equation}
\Delta=P^{4}+2P^{2}\left(H^{2}+E^{2}\cos^{2}Q\right)+(H-E^{2}\cos^{2}Q)^{2}
\geq 0,
\end{equation}
and thus real eigenvalues are guaranteed. (However, $\Delta \geq
0$ may \emph{not} imply the weak energy conditions.)

\qquad Assuming the existence of real eigenvalues, the
corresponding natural orthonormal eigenvectors are furnished by:
\begin{eqnarray}
E^{i}_{(2)}&=&r^{-1}\delta^{i}_{(2)},
E^{i}_{(3)}=(r\sin\theta)^{-1}\delta^{i}_{(3)},  \; E^{1}_{(1)}
=\frac{1}{2} \nu_{(1)}\left[T^{4}_{\;4}-
T^{1}_{\;1}-\sqrt{\Delta}\right], \nonumber \\
E^{2}_{(1)}&=&E^{3}_{(1)} \equiv 0, \;
E^{4}_{(1)}=-\nu_{(1)}T^{4}_{\;1},
E^{1}_{(4)}=-\nu_{(4)}T^{1}_{\;4},\;E^{2}_{(4)}=E^{3}_{(4)}\equiv
0, \label{eq:orthvecs} \\
E^{4}_{(4)}&=&\frac{1}{2}\nu_{(4)}\left(T^{1}_{\;1}-T^{4}_{\;4}+\sqrt{\Delta}\right),
\nonumber \\
\left[\nu_{(1)}\right]^{2}&:=& \frac{2e^{-\alpha}}{\sqrt{\Delta}
\left[T^{1}_{\;1}-T^{4}_{\;4}+ \sqrt{\Delta}\right]}, \nonumber \\
\left[\nu_{(4)}\right]^{2}&:=& \frac{2e^{-\gamma}}{\sqrt{\Delta}
\left[T^{1}_{\;1}-T^{4}_{\;4}+ \sqrt{\Delta}\right]}. \nonumber
\end{eqnarray}

\qquad The decomposition of the spherically symmetric $T^{ij}$ in
terms of the real eigenvalues and eigenvectors is accomplished by
\begin{equation}
T^{ij}=\left[\lambda_{(1)}- \lambda_{(2)}\right]
E^{i}_{(1)}E^{j}_{(1)}+ \lambda_{(2)}g^{ij} +\left[\lambda_{(2)}
-\lambda_{(4)} \right]E^{i}_{(4)}E^{j}_{(4)} \label{eq:teigen} .
\end{equation}
It is worth noting that the above algebraic structure of the
stress-energy tensor is common to \emph{many} different physical
arenas. For example, the anisotropic fluid is specified by
\begin{eqnarray}
p_{\parl}&:=&\lambda_{(1)},\; p_{\perp}:=\lambda_{(2)} \equiv
\lambda_{(3)} =T^{2}_{\;2},\; \mu:=-\lambda_{(4)}, \nonumber \\
u^{i}&:=&E^{i}_{(4)},\; s^{i}:=E^{i}_{(1)}, \nonumber \\
T^{ij}&=&\left(\mu+
p_{\perp}\right)u^{i}u^{j}+p_{\perp}g^{ij}+\left(
p_{\parl}-p_{\perp}\right)s^{i}s^{j}. \label{eq:anisosetens}
\end{eqnarray}
The physical quantities are the energy density ($\mu$) the radial
pressure ($p_{\parl}$) and the angular or transverse pressures
($p_{\perp}$). Anisotropic fluid models have received much
attention mainly in the arenas of stellar structure theory, black
holes and cosmology \cite{ref:eight}, \cite{ref:twelve},
\cite{ref:thirteen}, \cite{ref:otheraniso1}. Note that the
nomenclature ``anisotropic fluid'' is misleading. The
stress-energy tensor in (\ref{eq:anisosetens}) actually
represents a fluid which is \emph{not necessarily} isotropic.

\qquad In case
$\lambda_{(1)}\equiv\lambda_{(2)}\equiv\lambda_{(3)}$ (or
$p_{\parl}=p_{\perp}=:p$), the equation (\ref{eq:anisosetens})
yields the well known perfect fluid stress-energy tensor:
\begin{equation}
T^{ij}=\left(\mu+ p\right)u^{i}u^{j}+p g^{ij}.
\label{eq:perffluidsetens}
\end{equation}
This equation implies, by equation (\ref{eq:pressure}) and
(\ref{eq:eigvals}), the \emph{isotropy equation}:
\begin{eqnarray}
\sqrt{(T^{1}_{\;1}-T^{4}_{\;4})^{2}
-e^{2\alpha-h-\chi}(T^{1}_{\;4})^{2}}&=&rT^{1}_{\;1,1}+ \left[1
-\frac{r}{2} \alpha_{,1} + \frac{\kappa r^{2}}{2}e^{\alpha}
\left(T^{1}_{\;1}-T^{4}_{\;4}\right) \right]
\left(T^{1}_{\;1}-T^{4}_{\;4}\right) \nonumber \\
&&-e^{-(h+\chi)/2}
\left[e^{2\alpha-(h+\chi)/2}T^{1}_{\;4}\right]_{,4}.
\label{eq:isoeqn}
\end{eqnarray}
It is a formidable equation to solve in general (see
\cite{ref:visser} for detailed considerations of the static case.)

\qquad In case the spatial eigenvalues are identically zero, the
stress-energy tensor in (\ref{eq:anisosetens}) reduces to that of
an incoherent dust.

\qquad In case we identify $p=\lambda_{(2)}\equiv \lambda_{(3)},
\; \mu=-\lambda_{(4)}, \;\alpha:=\lambda_{(1)}-\lambda_{(2)}$,
the stress-energy tensor is,
\begin{equation}
T^{ij}=(\mu+p)u^{i}u^{j} +pg^{ij}+\alpha s^{i}s^{j}.
\label{eq:tachdustsetens}
\end{equation}
The above $T^{ij}$ is due to a perfect fluid plus a tachyonic
(space-like) dust. Such a stress-energy tensor has been
considered in a cosmological model \cite{ref:thirteen} where the
dust contributes to the dark matter or dark energy component of
the universe.

\qquad Now we are in a position to state and prove the main
theorems of this section involving anisotropic fluids.

{\theorem Let the spherically symmetric interior equations
(\ref{eq:einstone}-\ref{eq:einstfour}) and the conservation
equations (\ref{eq:cons1}), (\ref{eq:cons2}) hold in the
coordinate convex domain $D$ defined by (\ref{eq:domain}).
Moreover, let the stress-energy tensor $T^{i}_{\;j}$ be that of
an anisotropic fluid given by (\ref{eq:anisosetens}). Also, let
the physical conditions $T^{4}_{\;4} \leq 0$ and
$T^{1}_{\;1}-T^{4}_{\;4} \geq 0$ be satisfied in $D$. Then, the
most general solutions of all the equations and inequalities are
furnished by the following:
\begin{eqnarray}
0< q_{1} <1&,&  0 < q_{2}<1, \label{eq:slacksolution} \\
2M(r,t)&:=&\kappa q_{1} \left\{F^{2}(q_{2}t) +\int_{0^{+}}^{r}
E^{2}(x,q_{2}t)x^{2}\,dx\right\} >0, \nonumber \\
e^{-\alf}&=&1-\frac{2M(r,t)}{r}, \nonumber \\
e^{\gam}&=&e^{-\alf}\exp\left[h(t)+\chi(r,t)\right], \nonumber \\
\chi(r,t)&:=&\kappa q_{1}\int_{0^{+}} e^{\alpha(x,t)}
\left[E^{2}(x,q_{2}t)\cos^{2}Q(x,q_{2}t)+P^{2}(x,q_{2}t)
\right]x\,dx, \nonumber \\
\Delta(r,t)&=&(q_{1})^{2}\left[E^{2}\cos^{2}Q+P^{2}\right]^{2}
-\frac{4}{\kappa^{2}r^{4}}e^{\alpha-\gamma}
\left(M_{,4}\right)^{2} \geq 0, \nonumber \\
\frac{2}{q_{1}}\mu(r,t)&=& (q_{1})^{-1}\sqrt{\Delta} +E^{2}
\left[1+\sin^{2}Q\right]-P^{2} \geq 0, \nonumber \\
\frac{2}{q_{1}}p_{\parl}(r,t)&=& (q_{1})^{-1}\sqrt{\Delta} -E^{2}
\left[1+\sin^{2}Q\right]+P^{2} \geq 0, \nonumber
\end{eqnarray}
\begin{eqnarray}
\frac{p_{\perp}(r,t)}{q_{1}}&=&\frac{1}{2r} \left[r^{2} (P^{2}
-E^{2}\sin^{2}Q)\right]_{,1} \nonumber \\
&&+r\left\{\frac{1}{4}(E^{2}\cos^{2}Q+P^{2})\gamma_{,1}
-(kq_{1})^{-1} e^{-(\alpha+\gamma)/2}\left[ e^{(3\alpha-\gamma)/2}
M_{,4} \right]_{,4} \right\} \nonumber \\
u^{1}&=&-\frac{2\sqrt{2}}{\kappa e^{\gamma/2}r^{2} \Delta^{1/4}}
\frac{M_{,4}}{\sqrt{q_{1}\left(E^{2}\cos^{2}Q+P^{2}\right)+\sqrt{\Delta}}},
\;u^{2}=u^{3} \equiv 0, \nonumber \\
u^{4}&=&\frac{1}{\sqrt{2}e^{\gamma/2} \Delta^{1/4}}
\sqrt{q_{1}\left(E^{2}\cos^{2}Q+P^{2}\right)+\sqrt{\Delta}} > 0,  \\
s^{1}&=&\mp\frac{1}{\sqrt{2}e^{\alpha/2} \Delta^{1/4}}
\sqrt{q_{1}\left(E^{2}\cos^{2}Q+P^{2}\right)+\sqrt{\Delta}},\;
s^{2}=s^{3}\equiv 0, \nonumber \\
s^{4}&=&\mp \frac{2\sqrt{2}e^{\alpha/2-\gamma} M_{,4}}{\kappa
r^{2}
\Delta^{1/4}\sqrt{q_{1}\left(E^{2}\cos^{2}Q+P^{2}\right)+\sqrt{\Delta}}}
. \nonumber
\end{eqnarray}
Here, the functions $F(q_{2}t),\;h(t), E(r,q_{2}t)$ (not
identically zero) are of at least class $C^{3}$ in $D$. Aside from
these restrictions, the functions are arbitrary.}

\qquad For proof of the above theorem we used the equations
(\ref{eq:ealpha}-\ref{eq:pressure}), (\ref{eq:2M} -
\ref{eq:chi}), (\ref{eq:slack1}-\ref{eq:slack4}),
(\ref{eq:eigvals}), (\ref{eq:orthvecs}), (\ref{eq:teigen}) and
(\ref{eq:anisosetens}). The two parameters in
(\ref{eq:slacksolution}) may appear to be superfluous. However,
note that in the limit $q_{1}\rightarrow 0_{+}$, the solutions in
(\ref{eq:slacksolution}) yield the flat space metric. Moreover,
for $0<q_{1}<1$ and and $\lim q_{2} \rightarrow 0_{+}$, the
metric goes over to a static one. Furthermore, sufficiently small
positive values of $q_{1}$ and $q_{2}$ facilitate satisfaction of
the complicated inequalities $1-\frac{2M}{r} >0$ and $\Delta > 0$.

\qquad We consider here a specific example of an exotic black
hole. Consider the following:
\begin{eqnarray}
F(q_{2}t)&\equiv& 0, \;
E^{2}(r,q_{2}t):=\frac{jr^{j-3}}{(1-q_{2}t)^{j}}, \; 3 \leq j.
\nonumber \\
2M(r,t)&=&\kappa q_{1}\left[\frac{r}{1-q_{2}t}\right]^{j},\;
r=B(t):=\left(\frac{2m}{\kappa q_{1}}\right)^{1/j}
\left(1-q_{2}t\right), \nonumber \\
e^{-\alf}&=&\left[1-\frac{\kappa q_{1}r^{j-1}}{(1-q_{2}
t)^{j}}\right],
\label{eq:specexamp1} \\
T^{1}_{\;1}(r,q_{2} t) &:=&k^{-2}q_{1} E^{2}(r,q_{2} t), \;
\sqrt{3}
\leq k, \nonumber \\
\chi(r,q_{2}t)&=& -(1+k^{-2})\ln \left[1-\frac{\kappa
q_{1}r^{j-1}}{(1-q_{2}
t)^{j}}\right], \nonumber \\
e^{\gam}&=&\left[1-\frac{\kappa q_{1}r^{j-1}}{(1-q_{2}
t)^{j}}\right]^{-1/k^{2}}e^{h(t)}. \nonumber
\end{eqnarray}
The above describes a mathematically rigorous collapse model for
an anisotropic fluid black hole \cite{ref:eight}.

\qquad Now we shall consider the junction conditions for the
solutions given in (\ref{eq:specexamp1}). We state and prove the
following corollary to the preceding theorem.

{\cor Let the conditions stated in the previous theorem be valid
in $D$ with $E(r,q_{2} t) \neq 0$. Moreover, let both Synge's
junction conditions $T^{ij}M_{,j|\partial D}\equiv 0$ and the ISLD
junctions conditions $\left[K_{ij}\right]_{\partial D}=0$ hold on
$\partial D$. Then, either,
\begin{subequations}
\begin{align}
\sin^{2}Q(r,q_{2}t)\,E^{2}(r,q_{2} t) =& P^{2}(r, q_{2}t) +
\left[B(t)-r\right]^{\nu^{2}} N(r,q_{2}t) \geq 0, \nu^{2} \in
\left\{1,2\right\}\cup[3,\infty) \nonumber \\
\mbox{and}& \nonumber \\
F^{2}(q_{2}t)=&(c_{0})^{2} + \int_{t}^{t_{2}}d\tau \left\{
\int_{0_{+}}^{B(t)} x^{2} \frac{\partial}{\partial \tau}
\left[E^{2}(x,q_{2}\tau) \right]\,dx\right\} \geq 0, \label{eq:cor1}\\
\mbox {or else}& \nonumber \\
F^{2}(q_{2}t)=&f^{2}(q_{2}t) + \int_{t}^{t_{2}}d\tau \left\{
\int_{0_{+}}^{B(t)} x^{2} \frac{\partial}{\partial \tau}
\left[E^{2}(x,q_{2}\tau) \right]\,dx\right\}, \label{eq:cor2} \\
E^{2}(r,q_{2}t)=&\frac{P^{2}}{{\rm
\mbox{\textrm{sech}}}^{2}\left[R(r,q_{2}t) \right]
+\sin^{2}Q} >0, \;\dot{f}(q_{2}t) \neq 0, \nonumber \\
\mbox{and}& \nonumber \\
e^{h(t)}=&4\left\{(1-\frac{2m}{r})^{-2}e^{-\chi(r,t)}
\left[\frac{M_{,4}\cosh(R(r,q_{2}t))}{\kappa
q_{1}r^{2}E^{2}(r,q_{2}t)} \right]^{2} \right\}_{|r=B(t)} >0.
\nonumber
\end{align}
\end{subequations}
Here, $N(r,q_{2}t),\;R(r,q_{2}t)$ and $f(q_{2}t)$ are functions
of at least class $C^{3}$ in $D$ but otherwise arbitrary.}
\\
\emph{Proof}: By the equations (\ref{eq:slack3}) and
(\ref{eq:cor1}) it follows that
\begin{equation}
(q_{1})^{-1}T^{1}_{\;1}=-\left[B(t)-r\right]^{\nu^{2}}N(r,q_{2}t),
\; T^{1}_{\;1|r=B(t)}\equiv 0. \nonumber
\end{equation}
Furthermore,
\begin{equation}
\left[\frac{2M_{,4}}{\kappa
q_{1}}\right]_{|r=B(t)}=\frac{dF^{2}(q_{2}t)}{dt}
+\int_{0^{+}}^{B(t)} x^{2} \frac{\partial}{\partial t}
E^{2}(x,q_{2}t)\,dx \equiv 0. \nonumber
\end{equation}
Therefore, by (\ref{eq:bdotzero}) and (\ref{eq:israeli}), both
Synge's condition and the ISLD conditions are satisfied.

\qquad In the second case, by the equations in (\ref{eq:slack3})
and (\ref{eq:cor2}) it is deduced that
\begin{equation}
\kappa^{-1}T^{1}_{\;1}(r,q_{2}t)= P^{2}- E^{2}(r,q_{2}t)\sin^{2}Q
= E^{2}(r,q_{2}t)\mbox{sech}^{2}\left[R(r,q_{2}t)\right] > 0.
\end{equation}
Moreover, $M_{,4} \neq 0$ and
\begin{equation}
e^{h(t)}=\left(1-\frac{2m}{r}\right)^{-2}e^{-\chi(r,t)}
\left[\frac{(T^{1}_{\;4})^{2}}{-T^{4}_{\;4}T^{1}_{\;1}}\right]_{|r=B(t)}.
\nonumber
\end{equation}
Thus, both equations (\ref{eq:eht}) and (\ref{eq:israeliii}) are satisfied.
$\blacksquare$

\qquad We have previously proved \cite{ref:nine} that under the
two conditions $T^{1}_{\;4}\equiv 0$ and $T^{1}_{\;1,4}\equiv 0$,
the solutions of the equations (\ref{eq:einstone}-\ref{eq:cons2})
can be transformed into a static solution. This is the interior
version Birkhoff's theorem. We can investigate directly the
static limit of equations (\ref{eq:einstone}-\ref{eq:cons2}).
Under suitable assumptions, including $\frac{dM(r)}{dr}>0$, the
boundary, $\partial D$, of the spherical body is given by $r=b$,
a positive constant. Now, the general solution will be furnished
in the following  statement:

{\theorem Let the static version of the spherically symmetric
field equations  and one conservation law
(\ref{eq:einstone}-\ref{eq:cons2}) hold in the domain
$D:=\left\{(r,t): 0 < r <b,\; t_{1} < t< t_{2} \right\}$.
Moreover, let the stress-energy tensor be given by
(\ref{eq:anisosetens}), satisfying $\mu(r) >0,\; \mu(r)
+p_{\parl} >0$. If, in addition, both Synge's and the ISLD
junction conditions hold at $r=b$, then the general solutions of
the static equations are furnished by:
\begin{eqnarray}
0<q_{1}<1&,& b>0,\;c_{0}\in \mathbb{R},\;\nu^{2} \in
\left\{1,2\right\}\cup [3,\infty), \nonumber \\
2M(r)&=&\kappa \left[(c_{0})^{2}+
\int_{0}^{r}x^{2}\mu(x)\,dx \right] >0, \nonumber \\
e^{-\alpha(r)}&=& 1-\frac{2M(r)}{r}, \nonumber \\
\chi(r)&:=&\kappa \int_{0^{+}}^{r} e^{\alpha(x)}\left[\mu(x)+
p_{\parl}(x) \right]
x\,dx, \nonumber \\
e^{\gamma(r)}&=&e^{\chi(r) -\alpha(r)}, \\
\Delta(r)&=& \left[\mu(r)+p_{\parl}(r) \right]^{2} > 0,
\nonumber \\
\mu(r)&=&q_{1} E^{2}(r) >0,\; p_{\parl}=q_{1} \left[ P^{2}(r)
-\left(\sin^{2}Q(r)\right) E^{2}(r) \right], \nonumber \\
p_{\perp}&=&\frac{1}{2r} \left\{r^{2}p_{\parl\, ,1} +\frac{r}{4}
p_{\parl}\gamma_{,1} \right\} , \nonumber \\
u^{i}&=&e^{-\gamma(r)/2}\,\delta^{i}_{(4)}, \; s^{i}=\pm
e^{\alpha(r)/2}\delta^{i}_{(1)}, \nonumber \\
(\sin^{2}Q(r))E^{2}(r)&:=& P^{2}(r) +(b-r)^{\nu^{2}}N(r) >0.
\nonumber
\end{eqnarray}
Here, $E(r)$, $P(r)$, $Q(r)$ and $N(r)$ are functions of at least
class $C^{3}$. Moreover, these functions and the parameters
$c_{0}$ and $\nu$ are arbitrary
save for the restrictions imposed above.} \\
Proof follows from (\ref{eq:slacksolution}) and (\ref{eq:cor1}).
Note that to avoid a singularity at $r=0$ (if this is included in
the domain), the constant $c_{0}$ should be set equal to zero so
that $\lim_{r\rightarrow 0} M(r)=0$.

\qquad An illustrative example will be provided in the following:
\begin{eqnarray}
0<q_{1}<1,&\;& b>0,\; c_{0}=0, \nonumber \\
E^{2}(r):=3,&\;& 2M(r)=\kappa q_{1}r^{3}, \nonumber \\
3\sin^{2}Q(r)&:=&P^{2}(r)-3\kappa q_{1}(b-r) \times \nonumber \\
&&\times \left\{\frac{b+r}{ \left[3\sqrt{1-\kappa
q_{1}b^{2}}-\sqrt{1-\kappa q_{1}r^{2}}\right]
\left[\sqrt{1-\kappa q_{1}r^{2}}+\sqrt{1-\kappa
q_{1}b^{2}}\right]}\right\} >0, \nonumber \\
e^{\alpha(r)}&=&1-\kappa q_{1} r^{2}, \nonumber \\
e^{\gamma(r)}&=& \left[\frac{3\sqrt{1-\kappa q_{1}b^{2}}
-\sqrt{1-\kappa q_{1}r^{2}}}{3\sqrt{1-\kappa q_{1} b^{2}}-1}
\right]^{2}, \nonumber \\
\mu(r)&=&3q_{1},\;p_{\parl}\equiv p_{\perp}=:p(r)=3q_{1}
\left[\frac{\sqrt{1-\kappa q_{1}r^{2}} -\sqrt{1-\kappa
q_{1}b^{2}}}{3\sqrt{1-\kappa q_{1} b^{2}}-\sqrt{1-\kappa
q_{1}r^{2}}} \right], \nonumber \\
p(b)&=&0.
\end{eqnarray}
The above obviously yields the well known interior Schwarzschild
constant density solution.

\qquad One final example which illustrates the use of this scheme
is that of the inner layers of a static neutron star
\cite{ref:weinberg}. In this case, the energy density,
$\mu=T_{(4)(4)}=-T^{4}_{\;4}$ is known from the quantum mechanics
of degenerate Fermions. As well, there is the ultra-relativistic
fluid equation of state, which should be valid in the inner
layers of the star. We summarize as follows:
\begin{equation}
\mu(k_{F})=\frac{8\pi^{2}}{h^{3}}\int_{0}^{k_{F}}k^{2}(k^{2}+m_{n}^{2})^{1/2}\,dk
\nonumber
\end{equation}
Here $h$ is Planck's constant, $k_{F}$ is the Fermi-momentum and
$m_{n}$ is the neutron mass. Since the extreme relativistic limit
is employed, the mass terms may be neglected compared to the
Fermi momentum so that a pressure calculation gives:
\begin{equation}
p_{\parl}\equiv p_{\perp}=\frac{1}{3}\mu. \label{eq:neuteos}
\end{equation}
Also, $\mu(r)$ is obtained by utilizing (\ref{eq:neuteos}) along
with the linear combination
$G^{1}_{\;1}-G^{4}_{\;4}=\frac{2}{3}\kappa\mu(r)$. Now the
isotropy equation (\ref{eq:isoeqn}) of the general solution
reads, in terms of $\mu(r)$,
\begin{equation}
\mu(r)_{,r}=-\frac{4}{r^{2}}\mu(r)
\left[1-\frac{2M(r)}{r}\right]^{-1} \left[\frac{\kappa \mu(r)
r^{3}}{6}+M(r)\right] , \label{eq:neutmueqn}
\end{equation}
Noting relation (\ref{eq:massderiv1}) and assuming a series
solution in $r$ we arrive at the following:
\begin{equation}
M(r)=\frac{3}{14}r,\; \mu(r)=\frac{3}{7\kappa r^{2}}, \nonumber
\end{equation}
which, from (\ref{eq:ealpha}) and (\ref{eq:egamma}) yields:
\begin{equation}
e^{\gamma}=\frac{4}{7}\frac{r}{r_{0}}, \;e^{\alpha}=\frac{7}{4}
\end{equation}
with $r_{0}$ a constant. The neighborhood about $r=0$ is excised
as the singularity at this point is due to the ultra-relativistic
approximation. Also, this solution is not valid for outer layers
of the star where deviations from the ultra-relativistic case are
significant.

\section{Exotic spherically symmetric solutions}
\qquad The exact solutions in (\ref{eq:2M} - \ref{eq:chi}) can be
generalized by abandoning the weak energy conditions to express:
\begin{subequations}
\begin{align}
2\kappa^{-1}M(r,t)=&f^{2}(t)-\int_{0^{+}}^{r}
T^{4}_{\;4}(x,t)x^{2}\,dx, \label{eq:exmass} \\
e^{-\alf}=&1-\frac{2M(r,t)}{r} > 0, \label{eq:exetoalpha}\\
e^{\gam}=&\left[1-\frac{2M(r,t)}{r}\right]e^{\chi(r,t)}H(t), \label{eq:exetogamma} \\
\chi(r,t):=&\kappa\int_{0^{+}}^{r}e^{\alpha(x,t)}\left[
T^{1}_{\;1}(x,t)- T^{4}_{\;4}(x,t)\right] x\, dx, \label{eq:exchi} \\
H(t) \neq 0. \label{eq:exH}
\end{align}
\end{subequations}

\qquad There are many situations when the solutions to these
equations may prove to be ``exotic'' in some sense. For example,
the equations (\ref{eq:exetoalpha}-\ref{eq:exH}) reveal that the
condition \texttt{signature}$[g_{ij}]=+2$ may {\em not} be
preserved everywhere. A simple example may be considered in the
exact vacuum metric given by:
\begin{equation}
ds^{2}=\left(1-\frac{2m}{r}\right)^{-1}\,dr^{2}
+r^{2}\left(d\theta^{2} +\sin^{2}\theta\,d\phi^{2} \right)
+\left(1-\frac{2m}{r}\right)t^{3}\, dt^{2} .
\label{eq:signchngline}
\end{equation}
In case $t<0$, the metric is obviously transformable to the
Schwarzschild solution. However, for $t>0$ the line element
(\ref{eq:signchngline}) yields the spherically symmetric vacuum
gravitational instanton solution. In (\ref{eq:signchngline}),
$\lim_{t\rightarrow 0}g_{44}(r,t)=0$, indicating the existence of
a horizon. All the null rays from the Schwarzschild universe
suddenly halt on such a horizon. It may be called the
\emph{instanton horizon}. Signature changing metrics in general
relativity have been studied in \cite{ref:eight},
\cite{ref:othersigchng}. The most general spherically symmetric
instanton solution in curvature coordinates is furnished by the
equations (\ref{eq:exmass} - \ref{eq:exH})with the choice
$H(t)=-e^{h(t)} <0$.

\qquad Next, consider spherically symmetric $T$-domain solutions
(\cite{ref:six} and references therein). The metric is locally
expressible as:
\begin{align}
ds^{2}=&e^{\lambda(T,R)}\,dR^{2} + T^{2}\left(d\theta^{2}
+\sin^{2}\theta\, d\phi^{2} \right) -e^{\nu(T,R)}\,dT^{2}; \label{eq:tdommet} \\
D_{T}:=&\left\{(T,R):\;T_{1} < T < T_{2}, \; R_{1}<R<R_{2}
\right\}. \nonumber
\end{align}

Einstein's field equations $G^{i}_{\;j} +\kappa
\Theta^{i}_{\;j}=0$ can be solved with the metric
(\ref{eq:tdommet}). The general solution (``dual'' to the
solutions in (\ref{eq:ealpha} - \ref{eq:pressure})), is furnished
by:
\begin{subequations}
\begin{align}
 e^{-\nu(T,R)}=&\frac{1}{T} \left[\sigma(R) -\kappa
\int_{T_{0}}^{T} (T^{\prime})^{2} \Theta^{1}_{\;1}(T^{\prime},R)
\; dT^{\prime} \right] -1 =:\frac{2\Xi(T,R)}{T}-1 > 0\, , \label{eq:tetonu} \\
e^{\lambda(T,R)}=&\left[ \frac{2\Xi(T,R)}{T}-1 \right] \exp
\left\{\beta(R) +\kappa \int_{T_{0}}^{T} e^{\nu(T^{\prime},R)}
\left[
\Theta^{4}_{\;4}(T^{\prime},R)-\Theta^{1}_{\;1}(T^{\prime},R)
\right] T^{\prime} \, dT^{\prime}\right\},  \\
\Theta^{4}_{\;1}(T,R):=&\frac{2}{\kappa T^{2}}
\left[\Xi(T,R)\right]_{,1}, \\
\Theta^{2}_{\;2} \equiv \Theta^{3}_{\;3}=& \frac{T}{2}
\left[\Theta^{4}_{\;4,4} +\Theta^{1}_{\;4,1}\right] +
\left[1+\frac{T}{4} \lambda_{,4} \right] \Theta^{4}_{\;4} +
\frac{T}{4} \left(\lambda+\nu\right)_{,1} \Theta^{4}_{\;1}
-\frac{T}{2} \alpha_{,4} \Theta^{1}_{\;1},  \\
\mbox{all other }&\Theta^{i}_{\;j} \mbox{'s} \equiv 0. \nonumber
\end{align}
\end{subequations}
Here, the functions $\sigma(R)$ and $\beta(R)$ are of class
$C^{3}$ but otherwise arbitrary. The ``total tension'' function,
$\Xi(T,R)$ is generated by the tension density since, in the
$T$-domain, it is  $\Theta^{1}_{\;1}$ which appears in
(\ref{eq:tetonu}). This class of solutions includes eternal black
hole solutions.

\qquad Another special case of $T$-domain solutions occurs
whenever the stress-energy tensor matrix $\left[\Theta^{i}_{\;j}
\right]$ admits complex eigenvalues. The algebraic criterion of
such occurrence is provided by the strict inequality:
\begin{equation}
\Delta^{\sharp}:=\left(\Theta^{4}_{\;4}- \Theta^{1}_{\;1}
\right)^{2} +4\Theta^{1}_{\;4}\Theta^{4}_{\;1} <0.
\end{equation}
As an example, the following $\Theta^{i}_{\;j}$ has appeared in
the late stages of  gravitational collapse studies
\cite{ref:eight}:
\begin{align}
0<q_{1} <1, \;& \;0 <q_{2} <1, \; 3\leq j, \; \sqrt{3} \leq k,
\nonumber \\
\Theta^{1}_{\;1}(T,R)=&- \frac{jq_{1}
T^{j-3}}{\left(1-q_{2}R\right)^{j}} < 0, \nonumber \\
\Theta^{4}_{\;4}(T,R) =&k^{-2}j q_{1}
\frac{T^{j-3}}{(1-q_{2}R)^{j}} >0, \\
\Theta^{4}_{\;1}(T,R)=&\frac{jq_{1}q_{2}T^{j-2}}{\left(1-q_{2}R
\right)^{j}} > 0. \nonumber
\end{align}

\qquad Finally, we shall consider the spherically symmetric field
equations in an arbitrary $D$-dimensional manifold (with $D\geq
3$) \cite{ref:nine}. There has been much study on the possibility
of extra dimensions in light of superstring theories. In the low
energy sector, many of these theories reproduce a higher
dimensional general relativity in which, above some energy scale,
all dimensions may be considered non-compact. These higher
dimensional field equations may therefore have relevance in these
theories.

\qquad The metric in curvature coordinates is provided by:
\begin{equation}
ds^{2}=e^{\alf}\,dr^{2} +r^{2} \,d\Omega^{2}_{(D-2)}
-e^{\gam}\,dt^{2}\; ,
\end{equation}
with
\begin{align}
d\Omega^{2}_{(D-2)}=&\left[d\theta_{(0)}^{2}+\sum_{n=1}^{D-3}d\theta_{(n)}^{2}
\left(\prod_{m=1}^{n}\sin^{2}\theta_{(m-1)}\right)\right].
\nonumber \\
\tilde{D}:=&\left\{(r,\;\theta_{(0)}, \ldots, \theta_{(D-3)}, \;t)
\in \mathbb{R}^{D}:\;\; t_{1} < t < t_{2},\; 0 < r_{1} < r <
r_{2},
\right. \nonumber \\
& \left. \; 0 < \theta_{(0)},\ldots,\theta_{(D-4)} < \pi, \; 0
\leq \theta_{(D-3)} < 2\pi \right\}.
\end{align}
The $D$-dimensional field equations and conservation laws read:
\begin{subequations}
\begin{align}
\Epsilon^{1}_{\;1} =& \frac{D-2}{2
r^{2}}\left[(D-3)\left(1-e^{-\lam}\right) -r
e^{-\lam}\nw_{,1}\right]+ \kappa T^{1}_{\; 1}=0\;,  \label{eq:einstrr} \\
\Epsilon^{2}_{\;2} =& -\frac{e^{-\nw}}{4}\left[\nw_{,4}\lam_{,4} -
\left(\lam_{,4}\right)^{2} - 2\lam_{,44}\right]
 \left.- \frac{e^{-\lam}}{4}\left[2 \nw_{,11} +
\left(\nw_{,1}\right)^{2} +
\frac{2(D-3)}{r}\left(\nw-\lam\right)_{,1} \right.\right. \nonumber \\
 &\left. -\nw_{,1}\lam_{,1} +\frac{2}{r^{2}}(D-3)(D-4)\right]
+\frac{2(D-3)(D-4)}{r^{2}} + \kappa T^{2}_{\; 2} =0 \label{eq:einstthetatheta}. \\
\Epsilon^{4}_{\;1} =& \frac{D-2}{2r}e^{-\nw}\lam_{,4} + \kappa
T^{4}_{\; 1}=0\;,\;\; \Epsilon^{\theta_{n}}_{\;\theta_{n}}\equiv
\Epsilon^{2}_{\;2},
\label{eq:einsttr} \\
\Epsilon^{4}_{\;4} =&\frac{D-2}{2
r^{2}}\left[(D-3)\left(1-e^{-\lam}\right) +r
e^{-\lam} \lam_{,1}\right]+ \kappa T^{4}_{\;4} =0\;, \label{eq:einstt} \\
\Tau^{1}=&T^{1}_{\;1,1}+T^{4}_{\;1,4}+ \left[\frac{1}{2}\nw_{,1}
+\frac{D-2}{r}\right] T^{1}_{\; 1}+
\frac{1}{2}\left(\nw+\lam\right)_{,4} T^{4}_{\; 1}
-\left[\frac{1}{2}\nw_{,1}T^{4}_{\;4}+
\frac{D-2}{r}T^{2}_{\;2}\right]=0,
\label{eq:Dcons1} \\
\Tau^{4}=&T^{4}_{\;4,4} + T^{1}_{\;4,1} +\frac{1}{2}\lam_{,4}
\left(T^{4}_{\;4}- T^{1}_{\; 1}\right)+\frac{1}{2} T^{1}_{\; 4}
\left[\left(\lam + \nw \right)_{,1} + \frac{2(D-2)}{r}\right] =
0, \label{eq:Dcons2}
\end{align}
\end{subequations}

The general solution of the Einstein field equations and
conservation equations furnished utilizing the scheme in this
paper is:
\begin{subequations}
\begin{align}
e^{-\alf}=&1+\frac{2\kappa}{(D-2)r^{D-3}}\int_{r_{1}}^{r}T^{4}_{\;4}(x,t)
x^{D-2}\, dx -\frac{\kappa f^{2}(t)}{r^{D-3}}
=:1-\frac{2M(r,t)}{r^{D-3}}, \\
e^{\gam}=&e^{-\alf}
\exp\left\{h(t)+\frac{2\kappa}{D-2}\int_{r_{1}}^{r}
\left[\frac{T^{1}_{\; 1}(x,t) - T^{4}_{\;
4}(x,t)}{x^{D-3}-2M(x,t)}\right]x^{D-2} \, dx \right\}.
\end{align}
\end{subequations}
Again, the functions $f(t)$ and $h(t)$ are of class $C^{3}$ but
otherwise arbitrary.

\section{Concluding remarks}
In summary, the general solution to the spherically symmetric
Einstein field equations was provided in the case when the energy
density and parallel pressure are known. Both Synge's junction
conditions and the Israel-Sen-Lancsoz-Darmois junction conditions
have been studied and solved in general. The junction or boundary
is defined by the existence of a total interior mass which has
been rigorously proved in section 3. The weak energy conditions in
spherical symmetry for arbitrary boost were presented and solved
utilizing slack function methods. Specific matter models have
also been considered including the anisotropic fluid satisfying
both junction conditions, which includes the perfect fluid as a
special case. Finally, exotic extensions were considered.

\section*{Acknowledgements}
We would like to thank Tegai Sergei of the Department of Theoretical 
Physics at
Krasnoyarsk State University for bringing to our attention the
fact that case II of the ISLD junction conditions corresponds
to a non-time-like boundary. We are grateful for his careful reading of the manuscript.

\linespread{0.6}
\bibliographystyle{unsrt}

\end{document}